\documentclass[aps, prd, twocolumn, lengthcheck, superscriptaddress, 
nofootinbib]{revtex4-1}

\usepackage{epsfig}
\usepackage[usenames]{color}
\usepackage{graphicx}
\usepackage{amsmath}
\usepackage{epstopdf}

\newcommand\sect[1]{\emph{#1.}---}
\def\bi{\bibitem}

\def\la{\langle}\def\ra{\rangle}
\def\be{\begin{eqnarray}}\def\ee{\end{eqnarray}}
\def\lsim{\mathrel{\rlap{\lower3pt\hbox{\hskip1pt$\sim$}}
     \raise1pt\hbox{$<$}}} 
\def\gsim{\mathrel{\rlap{\lower3pt\hbox{\hskip1pt$\sim$}}
     \raise1pt\hbox{$>$}}} 
\def\del{\partial}

\def\calL{\cal L}

\allowdisplaybreaks

\begin{document}

\title{What ``Quenches"  ${g_A}$ in Nuclei ?}


\author{Mannque Rho}
\email{mannque.rho@ipht.fr}
\affiliation{Universit\'e Paris-Saclay, CNRS, CEA, Institut de Physique Th\'eorique, 91191, Gif-sur-Yvette, France }

\date{\today}

\begin{abstract}
 The answer is found in the way the hidden scale symmetry involving a  dilaton emerges in strong nuclear correlations in  nuclear matter.  It is suggested that the same mechanism is at the origin at higher densities of the  sound speed  converging in the core of massive compact stars to what could be called   ``pseudo-conformal sound speed" $v_s^2/c^2\approx 1/3$. A precision measurement of the superallowed Gamow-Teller transitions in the doubly magic nucleus $^{100}$Sn is suggested to confirm or falsify this prediction. It could also lead to the possible determination of genuine ``fundamental renormalization" of $g_A$ in nuclear medium. 
\end{abstract}

\maketitle

\sect{Introduction and results}
In a recent effort to formulate an effective field theory (EFT) that goes in density from $n\sim n_0\sim 0.16$ fm$^{-3}$ to densities relevant to the core of massive compact stars $n_{\rm NS}\sim (5-7)n_0$, a solution to the long-standing puzzle as to what makes the axial-vector coupling $g_A$ in the weak current ``quenched" to $\sim$ 1 from the free-space value 1.276, 
obtained somewhat heuristically a long time ago, was found to have a possibly deep origin in the way scale symmetry, hidden at nuclear matter density, intervenes  via strong nuclear correlations in superallowed Gamow-Teller transitions in nuclear medium. 
This is a totally new interpretation of an old observation which has remained completely obscure and ignored in nuclear physics for almost three decades. What I would like to describe here is what's {\it new} in it and why it can be of significance for future directions in nuclear theory. 

It should be stressed to start with that while the coupling constant $g_A$ in the weak current may actually be quenched by certain ``fundamental" renormalization (to be discussed in what follows),  what's referred to in the literature as ``quenched $g_A$" is a misnomer. It will seen that it is not the coupling constant $g_A$, but the matrix element of the Gamow-Teller operator that is quenched. This point accounts for the quotation mark.

That in Landau Fermi-liquid theory of nuclear matter, a quasiparticle on the Fermi surface makes a Gamow-Teller (GT)  transition with zero momentum transfer with an {\it effective} $g_A$ (denoted $g_A^L$) very close to 1, was predicted a long time ago. It followed in certain approximations, considered reasonable then,  from a chiral Lagrangian implemented with BR scaling via dilaton condensate in the sliding vacuum with density~\cite{br91} applied to quantum Fermi-liquid structure of nuclear matter~\cite{mr91,FR}.   However it was totally unclear what the puzzling result meant with respect to, or had anything to do with, what's observed in finite nuclei, particularly in simple shell model~\cite{review-gA}, or whether it signals a fundamental aspect of strong interactions encoded in QCD~\cite{wilkinson}.  For this reason  the intriguing observation $g_A^L\approx 1$ obtained in quantum Fermi-liquid theory~\cite{mr91,FR} has been more or less totally ignored in the field. 

In this paper, I re-derive the ``quenched" $g_A^L$ in the EFT of the quantum Fermi-liquid structure of nuclear matter. Although there is nothing new in the numerical result,
 it sharpens the notion and role of hidden symmetries, i.e., hidden local symmetry (HLS) and hidden scale symmetry (HSS), so far unrecognized, implemented into the EFT.   It represents a generalized nuclear EFT, denoted G$n$EFT in what follows,  that while reproducing  the  EoS of nuclear matter as well as the standard chiral EFT (SChiEFT) does up to density $\sim 2n_0$, becomes  applicable to higher densities  relevant to compact stars, $\sim (5-7) n_0$. Formulated in terms of the renormalization-group (RG) approach to baryons on the Fermi surface~\cite{EFT-fermi-surface},   the $g_A^L$  associated with  the $\beta$ decay of the quasiparticle on the Fermi surface can be interpreted to be equivalent to the ``quenched $g_A$" in the ``extreme single particle shell-model" (ESPSM) description. This is argued to provide a direct connection to observables in nuclei. One cannot make such connection to  non-doubly-magic nuclear systems studied extensively~\cite{review-gA}. For this,  a new formulation of Fermi-liquid theory for nuclei, such as the Migdal theory~\cite{migdal}, will be required. I will not go into it. Instead here I will take what I deem as a specially favorable case that involves doubly magic shells in shell model for a heavy nucleus, say,  $^{100}$Sn~\cite{FGG,Henke,RIKEN}. As will be seen, this offers specific advantages, both theoretical and experimental, for my arguments.

Let me  first give the principal results  predicted in G$n$EFT. The ``quenched $g_A$" observed in nuclei denoted $g_A^\ast$ is given by
\be
g_A^\ast=q_{\rm ssb} g_A^L
\ee  
where $q_{\rm ssb}$ is the fundamental quenching factor associated with the breaking, both explicit and spontaneous, of scale symmetry (which will be defined below). From Eq.~(\ref{gALP}), one obtains, within 10\% in the density range  $1/2 \lsim n/n_0\lsim 1$, 
\be
g_A^L\approx 1 .\label{gAL}
\ee 
For reasons that will become clear,  this result does not directly apply to  transitions involving non-negligible momentum transfers such as for instance in neutrinoless double beta decays where the momentum transfer can be of O(100) MeV. It also does not apply  to first forbidden beta decay transitions. 

\sect{G$n$EFT dictated by hidden symmetries}
The basic ingredients of  the G$n$EFT exploited here -- more precisely defined below -- are summarized in the 13 Propositions listed in \cite{MR-GnEFT}. The G$n$EFT was constructed to primarily address compact-star matter (at $n_{\rm NS}\sim (5-7) n_0$) while being capable to  capture the physics of the EoS of normal nuclear matter (at $\lsim 2 n_0$). It invokes a few assumptions, but most of the ingredients are interrelated and logically connected\footnote{The Propositions of \cite{MR-GnEFT} absolutely relevant for the discussions will be referred to as ``Propositions" in footnotes in what follows.}. It has only a few essential parameters that are given justifications to the extent feasible. How the  G$n$EFT approach works out is detailed in \cite{MR-GnEFT} and updated in \cite{MR-MPLA}.  I will focus on the density regime $n\sim n_0$ where the ``quenched $g_A$" figures although higher densities leading to what's called dilaton-limit fixed point (DLFP) are relevant.   

Let me start with a succinct description on the structure of the G$n$EFT needed for the problem. The notations are the same as in \cite{MR-GnEFT}.

The basic EFT Lagrangian denoted ${\cal L}_{\psi\chi \rm HLS}$ with which the G$n$EFT  is formulated consists of the two hidden symmetries, HLS and HSS, implemented to the standard chiral Lagrangian of the baryons $\psi$ and pions $\pi$. The role of the hidden symmetries is to represent heavy degrees of freedom (HdFs) to enable one to  access higher energy/density scales than those of  SChiEFT. HLS is introduced  gauge-equivalently to the non-linear sigma model, which is the basis for SChiEFT, at the leading chiral order~\cite{HLS} and HSS is implemented in the ``genuine dilaton" (GD) scheme proposed by Crewther and Tunstall~\cite{GD,CT}. The vector mesons $V=(\rho,\omega)$ figure in HLS and the dilaton $\sigma_d$ (identified with $f_0(500)$) figures in the ``conformal compensator field" $\chi =\frac{1}{f_\chi} e^{\sigma_d/f_\chi}$. Both represent heavy degrees of freedom (HdFs) in the Lagrangian coupled with the light degrees of freedom $\pi$.  The nucleons $\psi=(p,n)$ of course figure in nuclear systems but in the guise of heavy-baryon formalism. Assumed in the G$n$EFT is that the vector mesons have what's identified as the ``vector manifestation" (VM)  fixed-point at which the $\rho$-meson mass $m_\rho$ goes to zero, an important element for compact-star properties (e.g., the pseudo-conformal sound speed)\footnote{HLS is considered to represent the gluon fields in QCD via Seiberg duality. This reflects the role of the VM fixed point.},  and that the dilaton $\sigma_d$ is associated with the ``infrared" (IR) fixed point at which both chiral and scale symmetries are realized (in the chiral limit) in the Nambu-Goldstone mode with $m_{\sigma_d}$ and $m_\pi$ going massless while $m_N$ and $m_V$ are massive.

Although the primary objective of the G$n$EFT~\cite{MR-GnEFT} anchored on the Lagrangian ${\cal L}_{\psi\chi \rm HLS}$ is to access the density regime of compact stars, it is indispensable to possess the mechanism to account for possible crossover from hadrons at low densities to quarks/gluons at densities $\gsim (2-4)n_0$.  The HdFs brought in by the hidden symmetries is to simulate this hadrons-quarks continuity. This requires that the G$n$EFT must correctly describe nuclear properties  near -- and slightly above -- $n_0$ in the presence of the HdFs as SChiEFT does without the HdFs.  Without invoking explicit QCD degrees of freedom, the hadrons-quarks continuity is found to be feasible by the topology change from skyrmions to half skyrmions in the mesonic Lagrangian ${\cal L}_{\chi \rm HLS}$ that involves what resembles ``pseudo-gap" phenomenon in high-T superconductivity.  What characterizes this changeover is that in the half-skyrmion phase, while there can be chiral density waves, the quark condensate vanishes when space averaged,  thus simulating chiral symmetry restoration with, however, non-vanishing pion decay constant\footnote{See ``Proposition" VI}.  This feature that manifests at the topology change density  $n_{1/2}\sim 3n_0$  is not explicitly visible at nuclear matter density but could intervene in certain properties of normal nuclear matter as emergent (as in the $g_A^\ast$ problem). The topology change, the cusp in the symmetry energy $E_{\rm sym}$ at $n_{1/2}$, is observed in the crystal formulation of dense baryonic matter. The cusp structure, present in the crystal simulation,  cannot however be incorporated directly into G$n$EFT. But it can be  embedded in the parameters of the Lagrangian ${\cal L}_{\psi\chi \rm HLS}$ with the HdFs smoothing the cusp singularity. This feature plays an important role in the nuclear tensor forces, in particular for spin-isospin channels such as the Gamow-Teller transitions\footnote{See ``Propositions" VII and VIII.}.

The most essential ingredient for the $g_A$ problem is the role of the dilaton representing the hidden scale symmetry.  What is particularly relevant in nuclear interactions is the  notion of ``genuine dilaton"~\cite{GD,CT}\footnote{Note that this GD  scenario in QCD for $N_f\sim 3$ is quite different from the conformal window scenario for $N_f\sim 8$ discussed for going beyond the Standard Model (BSM). This structure of the IR fixed point appears to be controversial in the BSM circle. But as argued in \cite{MR-GnEFT}, this controversy does not appear to crucially affect applications to nuclear phenomena. This notion of scale symmetry in QCD is also shared by the ``conformal dilaton" of \cite{DDZ}.}. The hidden scale symmetry is incorporated into  G$n$EFT via the conformal compensator field $\chi$ (of mass dimension 1 and scale dimension 1) put into the baryonic HLS Lagrangian ${\cal L}_{\psi \rm HLS}$\footnote{The leading scale-chiral-order terms are of the form of Eqs. (5.1) and (5.2) of \cite{MR-GnEFT}.}
\be
 {\cal L}^{{\rm G}n{\rm EFT}} = {\cal L}_{\psi\chi \rm HLS} + V_d (\chi, \psi, V, \pi).\label{GnL}
\ee
Here ${\cal L}_{\psi\chi \rm HLS}$ is the baryonic HLS Lagrangian, the action of which is made scale-invariant by the powers in $\chi/f_\chi$,  and the scale-symmetry explicit breaking (trace anomaly and quark-mass) terms are lumped into the dilaton potential $V_d$.  The vacuum breaks scale symmetry spontaneously as it does chiral symmetry, so applied to baryonic matter, the parameters of the Lagrangian pick up density-dependent condensates, thereby ``hiding" the symmetries. There can also be other density dependences that arise in matching to QCD via correlators at given scales, termed ``intrinsic density dependence" (IDD)\footnote{See Proposition IX.}. 

Now there are a variety of ways to address many-body problems with the Lagrangian (\ref{GnL}). 

One approach is to generalize SChiEFT to a perturbative version of G$n$EFT by incorporating power expansion in $ {\cal L}^{{\rm G}n{\rm EFT}}$ with the parameters fixed at an appropriate scale higher than that of the SChiEFT, implementing into the well-established chiral counting the power counting in scale symmetry.  
\be
m_\chi^2\sim (\alpha_s-\alpha_{sIR})\sim O(p^2)\sim O(\del^2)
\ee
where $\alpha_{sIR}$ is the fine structure constant $g_s^2/4\pi$ at the IR fixed point.  Unfortunately the presently available expansion scheme, worked out in \cite{LMR},  is  too unwieldy as it stands and has not been applied to nuclear matter. This version will not be considered below.

\sect{Fermi-liquid fixed-point approach for EW response functions}
The G$n$EFT that I use was suggested in \cite{mr91} which amounts to applying RG to interacting baryons on the Fermi surface~\cite{EFT-fermi-surface} anchored on the effective Lagrangian (\ref{GnL}) endowed with the vacuum change in medium.\footnote{This defines precisely the G$n$EFT employed in this paper.  See Proposition IX.}  One can think of this approach  belonging to the class of density functional theory \`a la Hohenberg-Kohn theorem, as relativistic mean field theory (RMFT) does, in strong interactions. What is needed for the $g_A$ problem is the very G$n$EFT formulated for compact-star physics but effective in the lower density regime $n\sim n_0$ with, however, certain constraints given at higher density dictated by the dilaton-limit fixed point (DLFP) taken into account.\footnote{Explained below. See Proposition III.}

The quantities we are interested in are the low-momentum transfer response functions to the electroweak (EW) fields. Consider in particular the quasiparticle sitting on the Fermi surface making $(\vec{q}, q_0)=(\vec{0}, \sim 0)$  transfer transition. How to calculate the response functions in Wilsonian-group flow approaches is detailed in Section 5 of \cite{MR-GnEFT}. The zero-momentum-transfer EW processes can be accurately calculated in the Fermi-liquid fixed point (FLFP) theory which amounts to the mean field approximation with the Lagrangian (\ref{GnL})\footnote{See Proposition X.}\label{X}. The calculation was done with certain drastic assumptions two and half decades ago in \cite{FR}. What comes out in the FLFP approximation,  remarkably,  confirms exactly what was obtained in  \cite{FR}. It is actually easy to understand how this comes about. What's involved is that the FLFP calculation satisfies in-medium Ward identities with both the EM and weak response functions,  encoding low-energy theorems that are correctly captured in the effective Lagrangian.

Before confronting the old results predicted in \cite{FR} in the light of the new developments, let me briefly remind the readers of what happens at high density approaching the IR fixed point. For this purpose, redefine the field 
\be
{\cal Z}=e^{i\vec{\tau}\cdot \vec{\pi}} f_\pi/f_\chi=s +i\vec{\tau}\cdot \vec{\pi}\label{Z}
\ee  
introducing the scalar field $s$ and letting 
${\rm Tr} {(\cal{Z}\cal{Z}^\dagger)} \to 0$ 
in ${\cal L}_{\psi\chi \rm HLS}$ to have the dilaton mass go to zero. This leads to the DLFP first discussed in \cite{DLFP}
\be
g_A^\ast\to 1, \ \ f_\chi^\ast\to f_\pi^\ast.\label{DLFP}
\ee
For the purpose of the problem concerned,  while the DLFP density is  relevant to the physics of compact stars,  what matters for the present discussion is that  $n_{\rm DLFP} > n_{1/2}$. Suppose that in the vicinity of the DLFP, the in-medium  Goldberger-Treiman relation for $\pi$  and its analog for $\chi$ hold as in the GD scenario in matter-free space~\cite{GD}. Then one can write 
\be
m_N^\ast=f_\chi^\ast g_{\chi NN}^\ast=f_\pi^\ast g_{\pi NN}^\ast/g_A^\ast
\ee
which leads, via (\ref{DLFP}),  to
\be
g_{\pi NN}^\ast= g_{\chi NN}^\ast.
\ee
 This is the relation that  one gets if $s$ in (\ref{Z}) is identified with $\sigma$, i.e., the fourth component of the chiral four-vector in the linear sigma model. This agrees with  the line of arguments developed by Yamawaki on how to go from non-linear sigma model to linear sigma model to scale-invariant model by dialing a single coupling constant $\lambda$~\cite{yamawaki}. Here the dialing is done by density.

The zero momentum transfer quasiparticle response function on the Fermi surface is calculated at the FL fixed point where $1/\bar{N}\to 0$ where $\bar{N}=k_F/(\Lambda_{\rm FS}-k_F)$ with $\Lambda_{\rm FS}$ being the cutoff on top of the Fermi surface~\cite{EFT-fermi-surface}. The single-decimation procedure corresponds to setting $\Lambda_{\rm FS}\to k_F$. The results of this calculation obtained  in \cite{FR} with extremely simple Lagrangian turned out to be reproduced in the mean field approximation of  the more realistic G$n$EFT. Among the quantities obtained in \cite{FR}, what's relevant to the quenched $g_A$ problem are the proton anomalous orbital gyromagnetic ratio $\delta g^p_l$ and what could be identified as  ``Fermi-liquid fixed-point  axial coupling constant" $g_A^L$
\be
\delta g_l^p &=& \frac 49\large(\Phi^{-1} - 1 -\frac 12\tilde{F}_1^\pi\large)^{-1}\label{deltagl}\\ 
g_A^L &=& g_A (1-\frac{\Phi}{\tilde{F}_1^\pi})^{-2}\label{gALP}.
\ee
where ${\tilde{F}_1^\pi}$ is the pionic contribution to the Landau parameter ${\tilde{F}_1}$ and $\Phi=f_\chi^\ast/f_\chi=f_\pi^\ast/f_\pi$. As mentioned, these relations correspond to Ward identities for the nuclear EW responses, the former for the EM and the latter for the weak current.  Remarkably they involve only two quantities, both of which are entirely given by chiral-scale symmetries. They are accurately calculable with the given chiral-scale Lagrangian, very well supported by experiments.  

The $\delta g_l^p$, a Landau fixed-point quantity, is an accurately  measurable quantity. The prediction (\ref{deltagl}) giving $\delta g^p_l= 0.22$ for $n\approx n_0$ agrees extremely well with the measured value in the Pb region $ \delta {g_l^p}^{\rm exp} =0.23\pm 0.03$ (quoted in \cite{FR}). Now  $g_A^L\approx 1$,  given in terms of the fixed-point quantities, is also a fixed-point quantity. What it represents physically was not clearly identified in \cite{FR}. It is here that the modern developments bring the clarification of what it represents and how it relates to the ``quenched $g_A^\ast$".

\sect{What is $g_A^L$ and how it can be connected to $g_A^\ast$?}

In  Fermi liquid theory for nuclear systems, there is a difference between $\delta g_l$ and $g_A$  in that the former is a static quantity involving the diagonal matrix element of the EM current whereas $g_A^\ast$ is involved in the transition matrix element of the axial current in nuclear systems with non-zero energy transfer. In particular, we are dealing here with  superallowed  Gamow-Teller transitions in the FLFP approximation with $|{\vec{q}}|/q_0\to 0 $ with $\vec{q} = 0$ and $q_0\gsim 0$. 

Now to precisely compare what's calculated with what's measured in experiments, one needs to zero-in on the relevant axial current. To the leading order in the scale-chiral power counting, the most general {\it in-medium} nucleon axial current given in the GD scheme~\cite{GD,CT} takes the form
\be
 J^{\pm}_{5\mu} = g_A q_{\rm ssb}\bar{\psi}\tau^{\pm}\gamma_\mu\gamma_5\psi +\cdots\label{axialcurrent} 
\ee 
where $\cdots$ represents polynomial terms in fluctuating $\chi^\prime$ field\footnote{They enter in higher-order terms in scale-chiral expansion that can be ignored in the $g_A$ problem.} and  
\be
q_{\rm ssb}=c_{\rm GT}+(1-c_{\rm GT})\Phi^{\beta^\prime}
\ee
where $c_{\rm GT}$ is in general a density-dependent constant that cannot at present be calculated by the theory  and $\beta^\prime$ is the derivative of the $\beta$ function with respect to $\alpha_s$, in principle  also density-dependent. The $\beta^\prime$ is not known for QCD with $N_f\leq 3$ even in the vacuum except that $\beta^\prime >0$ in the GD scheme. 

Note that $q_{\rm ssb}=1$ in the vacuum, so $q_{\rm ssb}\neq 1$ represents scale symmetry breaking  {\it anomaly-induced by nuclear medium.} In the matter-free vacuum, the axial current $J^{\pm}_{5\mu}$ is scale-invariant and the axial coupling is (fundamentally) non-renormalized. On the contrary,  the axial coupling can be renormalized in nuclear medium to $g_A^{\rm ren}=q_{\rm ssb} g_A$ if $q_{\rm ssb}\neq 1$.  A deep question is whether this {\it fundamentally renormalized} axial coupling intervenes in the puzzling issue of ``quenched $g_A$" in nuclei.  

It is worth stressing  that the above point, which could provide a highly non-trivial solution to the ``quenched $g_A$ problem," has been up to date totally unrecognized in nuclear physics community. 

In the formulation as defined, in the FLFP limit, the quasiparticle Gamow-Teller transition $I(0^+, T=0) \to F(0^-, T=1)$ for $q=0$ and $q_0\sim0$ is given by 
\be
M_{GT}=q_{\rm ssb}  g_A^L  (\tau^{\pm}\sigma )_{FI}\label{MGT} 
\ee
with $g_A^L$ given by Eq.~(\ref{gALP}). For the parameters $\bar{F}_1^\pi$ and $\Phi$ for density $n\sim n_0$, $g_A^L$ comes out to be $\approx 1$, Eq.~(\ref{gAL}), as announced. One can then identify
\be
g_A^\ast=q_{\rm ssb}g_A^L.\label{gA-ssb}
\ee
If $g_A^\ast\approx 1$ as indicated in the simple shell model for  the parent and daughter states~\cite{review-gA}, then this would imply that $q_{\rm ssb}\approx 1$ and hence the quenching is entirely encoded in nuclear correlations dictated by the emerging scale symmetry with no (significant) fundamental renormalization of the coupling constant $g_A$.

To appreciate what this means in terms of SChiEFT, one should note that the Fermi-liquid fixed-point limit is tantamount to taking into account all multi-body current operators in addition to one-body that enter in consistency with the chiral counting~\cite{br91}. For the Gamow-Teller operator, the $n$-body (for $n\geq 2$) terms are strongly suppressed in the chiral counting relative to the leading 1-body operator, therefore one is led to conclude  that only the one-body operator $\tau^\pm \sigma$ will figure non-negligibly in (\ref{MGT})\footnote{This was the argument to suggest~\cite{gA-MR} that the solution to the $g_A$ problem offered in \cite{firstprinciple} is unfounded.}. 

\sect{Connection of $g_A^L$ to experimental $g_A^\ast$}
In order to test the validity of $g_A^L$ -- that modulo $q_{\rm ssb}$, there is no bona-fide quenching of the coupling constant in $g_A^\ast=1$ -- it is necessary to see what the connection of the FLFP value of $g_A^L$ to the experimentally measured quantity $g_A^\ast$ is. Experimental results are analyzed almost entirely in shell model, so the question would be what is the shell-model quantity that has the physics closest to what's captured in the FLFP approximation. As already mentioned, the renormalization-group approach anchored on EFT is simplest -- and reliable -- when applied to the case where density is high and  $\bar{N}\to \infty$, which is precisely what the Fermi-liquid fixed approximation implies in G$n$EFT. This requires that  particle-hole  excitations on top of the ground state in the shell-model description should be strongly gapped. The best possible case that can be matched to the FLFP theory is therefore the experimentally feasible beta decay of the nucleus $^{100}$Sn. This nucleus has doubly magic closed 50-50 shells and particle-hole excitations are maximally gapped. . 

In order to make the arguments quantitatively accurate, the details of the observational (experimental) conditions will have to be scrutinized.  I will eschew them as they are too technically involved for me. (See  \cite{FGG,Henke} for details). I will simply assume that there is a single $(1^+, T=1)$ daughter state largely separated from higher excitations of the same quantum numbers in $^{100}$In to which the transition from the ground $(0^+, T=0)$ state in $^{100}$Sn is made.
I suggest that it is the ``extreme single-particle shell model" (ESPSM)  that can be best equated to the FLFP description at the $1/\bar{N}\to 0$ limit.   In terms of the ESPSM matrix element squared ${\cal B}^{\rm ESPSM}_{\rm GT}$, the experimental $g_A^\ast$ should be given by
\be
{g_A^\ast}^{\rm exp}=g_A \big({\cal B}^{\rm exp}_{\rm GT}/{\cal B}^{\rm ESPSM}_{\rm GT}\big)^{1/2}\label{gAexp}
\ee
with ${\cal B}^{\rm ESPSM}_{\rm GT}=16/9=17.78$~\cite{FGG}. Now the question is what the daughter state in shell model  is that best corresponds to the quasineutron to which the quasiproton makes the Gamow-Teller transition in the FLFP theory.  It should be a single dominant $(1^+, T=1)$ state as strongly gapped as possible from all higher excitations in ESPSM. This condition seems fairly well met in the doubly magic nuclear system. However in reality, there seem to be some complications in the identification of the daughter state involved. It seems that this makes the experimental analysis (\ref{gAexp}) rather involved~\cite{Henke,RIKEN} and the situation less clear.

I will take two extreme cases to illustrate the issues involved.

The first case is based on the GSI data~\cite{Henke}.  Let me take the case of confining to a single 
dominant daughter ($p_{g_{9/2}}^{-1}$-$n_{g_{7/2}}$)$^{1^+}$ state. Since the errors are big, I will use the central value. Roughly the range of the extracted experimental matrix element is 

\be
{\cal B}^{\rm exp,\rm GSI}_{\rm GT}= 9.1 - 10.
\ee
This leads to
\be
g_A^{\ast, \rm GSI}=0.91- 0.96.
\ee
I take this amply consistent with $g_A^\ast = 1$ and hence with $q_{\rm ssb} =1$. There is little, if any, fundamental renormalization and the ``quenching" captures the working of scale symmetry manifested in $g_A^L= 1$. This would be consistent with what's globally observed in light nuclei~\cite{review-gA,zuker}.

Now let me consider the totally different situation presented by the recent RIKEN analysis~\cite{RIKEN} which is claimed to be ``improved" with much smaller error bars.  I skip the involved details in the analysis and look at the matrix element quoted in the conclusion  
\be
{\cal B}^{\rm exp,\rm RIKEN}_{\rm GT}= 4.4^{+0.9}_{-0.7}.\label{Rikenexp}
\ee
 This gives the range 
\be
{g_A^\ast}^{\rm exp}\approx 0.6 - 0.7
\ee
which means by (\ref{gA-ssb}) and (\ref{gAL})
\be
q_{\rm ssb}\approx  0.6 - 0.7.
\ee
This would imply $g_A$ is {\it fundamentally quenched}  by as much as  40\% from the free-space value 1.276. One expects that $q_{\rm ssb}$ should be insensitive to density and hence will apply to all weak processes in nuclear medium. This would be a huge {\it fundamental} quenching of the coupling constant which if reconfirmed will have a dramatic impact on nuclear weak processes as a whole.

\sect{Conclusions}
It is argued, based on the Fermi-liquid fixed-point approach to nuclear matter ``mapped" to  the ESPSM approach in doubly magic nuclei, that  what makes the effective $g_A$ quenched to $g_A^\ast \approx 1$ -- modulo possible fundamental renormalization -- in nuclear matter is the hidden scale symmetry that is responsible for $g_A\to 1$ and $f_\pi\to f_\chi$ at high density approaching the dilaton limit fixed point as well as for the speed of sound in compact stars converging to $v_{pcs}^2/c^2\to 1/3$ in the core of the stars~\cite{MR-GnEFT,MR-MPLA}. It is neither the many-body currents~\cite{firstprinciple} nor  other ``exotic mechanisms" often invoked in the literature. 

At present, there is a serious discrepancy in experiments:  The most recent RIKEN data on $^{100}$Sn decay would indicate that there can be a substantial anomaly-driven {\it fundamental} quenching of $g_A$  induced in medium which has not been suspected up to date. If the RIKEN data and analysis were confirmed,  it could give hint on the role of $\beta^\prime$, one of the crucial quantities in QCD, thus far unavailable in the matte-free vacuum. 

Such a huge quenching of $g_A$, at present, is ruled out by other observables in nuclear beta decay observations. For instance, it is strongly at odds with what has been reliably established in the first-forbidden beta decays measured in the Pb region and predicted by G$n$EFT anchored on low-energy theorems~\cite{firstforbidden}.   What's perhaps more serious is that it would impact the neutrinoless double beta decay experiments in search for the BSM where Gamow-Teller matrix elements enter crucially in the analysis.  It is therefore of great importance in settling the important issue to make precision re-measurements and re-analyses of the superallowed Gamow-Teller decay in the doubly magic nucleus $^{100}$Sn.

\end{document}